\documentstyle[epsfig,subfigure,12pt,here]{article}

\protect
\protect
\newcommand{\lgm}{{\,\rm ln }}
\newcommand{\Break}{ \right. \nonumber \\ &{}& \left. }

\newcommand{\ice}[1]{\relax}

\newcommand{\prd}{\partial}
\newcommand{\beq}{\begin{equation}}
\newcommand{\eeq}{\end{equation}}
\newcommand{\api}{\frac{\alpha_s}{\pi}}

\newcommand{\ba}{\begin{array}} 
\newcommand{\ea}{\end{array}} 
\newcommand{\ds}{\displaystyle} 
\newcommand{\as}{\alpha_s}
\newcommand{\gm}{\gamma_m}
\newcommand{\G}{\Gamma}
\newcommand{\g}{\gamma}
\newcommand{\gaam}{\gamma^{AA}_m}

\newcommand{\gvvm}{\gamma^{VV}_m}
\newcommand{\gvvq}{\gamma^{VV}_q}

\newcommand{\gssq}{\gamma^{SS}_q}

\newcommand{\dmu}{\mu^2\frac{d}{d\mu^2}}
\newcommand{\msbar}{\overline{\mbox{MS}}}
\newcommand{\dsp}{\displaystyle}
\newcommand{\EQN}{\label}
\newcommand{\ovl}{\overline}

\begin{document}

\begin{titlepage}
\noindent

\begin{flushright}
\begin{tabular}{l}
  MPI/PhT/96-84\\
  hep-ph/9609202\\
  August   1996   
\end{tabular}
\end{flushright}

\protect\vspace*{.3cm}

\begin{center}
  \begin{Large}
  \begin{bf}
Quadratic  Mass Corrections of Order ${\cal O}(\alpha_s^3 m_q^2/s)$    
to the Decay Rate of $Z$- and $W$- Bosons  
  \\
  \end{bf}
  \end{Large}
  \vspace{1cm}
K.G.~Chetyrkin $^{a,b}$,
J.H.~K\"uhn$^{c}$,
\begin{itemize}
\item[$^a$]
 Institute for Nuclear Research,   
 Russian Academy of Sciences,   \\
 60th October Anniversary Prospect 7a 
 Moscow 117312, Russia 
\item[$^b$]{%
Max-Planck-Institut f\"ur Physik, Werner-Heisenberg-Institut, \\ 
F\"ohringer Ring 6, 80805 Munich, Germany}
\item[$^c$]
    Institut f\"ur Theoretische Teilchenphysik,
    Universit\"at Karlsruhe \\ 
    D-76128 Karlsruhe, Germany
\end{itemize}
\vspace{0.2cm}

  \vspace{0.5cm}
  {\bf Abstract}
\end{center}
\begin{quotation}
\noindent
We analytically compute  quadratic mass corrections of order 
${\cal O}(\alpha_s^3 m_q^2/s)$ 
to the absorptive part of the (non-diagonal) correlator of two axial vector 
currents. This allows us to find the correction of order 
${\cal O}(\alpha_s^3 m_q^2/M^2_W)$ to $\G(W \to \mbox{hadrons})$ as well as 
similar corrections to $\G(Z \to \mbox{hadrons})$. 
\end{quotation}
\vfill

\noindent
emails:
\\
chet@mppmu.mpg.de
\\
johann.kuehn@physik.uni-karlsruhe.de  \\ 

\vfill

\ice{
\noindent The complete postscript file of this
preprint, including figures, is available via anonymous ftp at
www-ttp.physik.uni-karlsruhe.de (129.13.102.139) as /ttp96-06/ttp96-06.ps 
or via www at http://www-ttp.physik.uni-karlsruhe.de/cgi-bin/preprints.
}
\end{titlepage}

\section{Introduction}
Precision measurements of the total and 
as well as
partial Z decay  rates have provided one  of the 
the most important and,  from  the theoretical viewpoint, clean
determination  of the strong coupling  constant $\alpha_s$
with a present value of $\alpha_s= 0.1202 \pm 0.0033$
\cite{Blondel}.  Theoretical ingredients were the knowledge
of QCD corrections to order $\alpha_s^3$ in the limit of massless
quarks plus charm and bottom quarks effects (see, e.g. \cite{review}
and references therein). These mass corrections which indeed are
relevant at the present level of accuracy have been calculated up to
the order $\alpha_s^3 m_q^2/s$ for the vector and $\alpha_s^2 m_q^2/s$
for the axial current induced decay  rate. In this short note the
prediction is extended to include $\alpha_s^3 m_q^2/s$ terms for the
(non-singlet part) of the axial current induced rate.  At the same
time results are obtained for the non-diagonal current correlator
with two different masses --  a case of relevance e.g. for the W
decay rate into charmed and bottom quarks. 
The same formulae can also be applied  to a subclass of
corrections which enter single top production in the Drell-Yan 
like reaction 
$q \ovl{q}\to  t \ovl{b}$ far above threshold.

The calculation is based on an approach introduced in
Refs.~\cite{ChetKuhn90,CheKueKwi92}. Knowledge of the polarization
function to order $\alpha_s^2$, the appropriate anomalous dimensions
at order $\alpha_s^3$, combined with the renormalization group
equation allows one to predict the corresponding logarithmic terms of
order $\alpha_s^3$ and hence the constant terms of the imaginary
part. The first of these ingredients has been available since some
time \cite{GorKatLarSur90,Chetyrkin93,Karl94,levan94} 
while the anomalous dimension can been obtained from Ref.~\cite{gssq}
in a straightforward way.

In this  short note  only the  theoretical framework and
the  analytical results are presented -- numerical studies will 
presented  elsewhere.

\section{Renormalization Group Analysis}
In analogy to the vector case, we take as a starting point the generic
vector/axial quark current correlator $\Pi^{V/A}_{\mu\nu}$ which is
defined by
\beq 
\ba{ll} 
\Pi^{V/A}_{\mu\nu}(q,m_u,m_d,m{},\mu,\as) & = \ds i \int dx e^{iqx} 
\langle 
T[\, j^{V/A}_{\mu}(x) (j^{V/A}_\nu)^{\dagger}(0)\, ] \rangle 
\\ &  = \ds g_{\mu\nu}  \Pi^{(1)}_{V/A}(Q^2) 
      +  q_{\mu}q_{\nu} 
  \Pi^{(2)}_{V/A}(Q^2))
{}.  
\ea 
\label{correlator}
\eeq
with $Q^2=-q^2$, $m^2_q = \sum_f m_f^2$ 
and  $j^{V/A}_{\mu} = \bar{q}\gamma_{\mu}(\gamma_5) q'$.
Here $q$ and $q'$ are just two (generically different) quarks with masses $m_u$
and  $m_d$ respectively. Note that the vector and axial correlators  
are related through 
\beq
\Pi^{A}_{\mu\nu}(q,m_u,m_d,m{},\mu,\as)
=
\Pi^{V}_{\mu\nu}(q,m_u,-m_d,m{},\mu,\as)
\label{VAidentity}
\eeq

The polarization function $\Pi^{(1)}_{V/A}$ and the spectral density
$R^{V/A}(s)$ which in turn governs the $Z$  and $W$ decays
rate obey the following dispersion relation 
\beq 
\dsp
\Pi_{A}^{(1)} (Q^2) = \frac{-1}{12\pi^2}
\int_{(m_u+m_d)^2}^{\infty}ds
\frac{s R^{V/A}(s,m_u,m_d,m{},\mu,\as)}{s+Q^2}
\ \ \  \;\;{\rm mod \;sub.} 
\label{dispersion.rel}
\eeq
Whereas $R^{V/A}$ as a physical quantity is invariant under 
renormalization group transformations,
the function 
$
\langle 
T[\, j^{V/A}_{\mu}(x) (j^{V/A}_\nu)^{\dagger}(0)\, ]
\rangle 
$ 
contains 
some non-integrable singularities in the vicinity of the point $x=0$.
These cannot be removed by standard quark mass and coupling constant
renormalizations, but must be subtracted independently.
As a result the relevant renormalization group equation 
assumes the form \cite{review}
\beq 
 \label{rgea}
\dmu \Pi^{V/A}_{\mu\nu} = (q_{\mu}q_{\nu}-g_{\mu\nu}q^2) \g^\pm_q(\as)
\frac{1}{16\pi^2} 
+ 
(m_u \mp m_d)^2 g_{\mu\nu} \g^\pm_m (\as) \frac{1}{16\pi^2},
\eeq
where 
\beq
\label{rgdef} \mu^2\frac{d}{d\mu^2} =
\mu^2\frac{\partial}{\partial\mu^2} + \pi\beta(\alpha_s)
\frac{\partial}{\partial \alpha_s} 
+\gm(\alpha_s)
\sum_f
 \bar{m_f}
\frac{\partial}{\partial \bar{m_f}} {}. 
 \eeq 
Here and
below the upper and lower signs give the results for vector and axial
vector correlators respectively.  
{}From the identity
(\ref{VAidentity}) we infer that both anomalous dimensions $\g^\pm_q$
and $\g^\pm_m$, being not dependent on any masses, also do not depend
on the sign. In what follows we will denote
\[
\g^\pm_q = \gvvq \ \ \ \mbox{and} \ \ \ \g^\pm_m = \gvvm \
{}.
\]
The $\beta$-function and the quark mass anomalous dimension $\gm$ are defined
in the usual way
\beq 
\dmu \left( \frac{\as(\mu)}{\pi} \right) = \as \beta(\as) \equiv
-\sum_{i\geq0}\beta_i\left(\api\right)^{i+1}, \eeq
\beq 
\dmu \bar{m}(\mu) =  \bar{m}(\mu)\gm(\as) \equiv 
-\bar{m}\sum_{i\geq0}\gm^i\left(\api\right)^{i+1}. \eeq
Their expansion coefficients up to order ${\cal O}(\as^3)$ are well known 
\cite{beta,Larin:betaQCD,gamma,Larin:massQCD} and read ($n_f$ is the number of quark flavours)
\beq
\ba{c}\dsp
\beta_0=\left(11-\frac{2}{3}n_f\right)/4,  \  \  
\beta_1=\left(102-\frac{38}{3}f\right)/16, \\ \dsp 
\beta_2=\left(\frac{2857}{2}-\frac{5033}{18}n_f+ 
\frac{325}{54}n_f^2\right)/64, 
\ea 
\label{beta3} 
\eeq 
\beq 
\ba{c}\dsp 
\g^0_m=1, \  \ \ \g^1_m=\left(\frac{202}{3}-\frac{20}{9}f\right)/16, 
\\ \dsp \g^2_m=\left(1249 - 
\left[\frac{2216}{27}+\frac{160}{3}\zeta(3)\right] 
n_f-\frac{140}{81}f^2\right)/64.  
\ea 
\label{anom.mass3} \eeq

Another useful  and closely related object is the correlator
of the (pseudo)scalar quark currents
\begin{equation}
\label{SP}
\Pi^{{\rm S/P}}(Q^2,m_u,m_d,m{},\mu,\as) = \int  
e^{iqx}\langle0|\; T\; [\,j^{{\rm S/P}}(x) (j^{{\rm S/P})\dagger}\,](0)
\;|0\rangle
{}\, .
\end{equation}
Scalar 
and  pseudoscalar
current correlators are also related in a simple
manner:
\begin{equation}
\Pi^{{\rm S}}(Q^2,m_u,m_d,m{},\mu) = \Pi^{{\rm P}}(Q^2,m_u,-m_d,m{},\mu)
{}.
\label{SPidentity}
\end{equation}
For vanishing quark masses
scalar and  pseudoscalar 
correlators  are, therefore, identical:
$\Pi^{{\rm S}}= \Pi^{{\rm P}}$
and meet the following
 RG  equation 
\beq 
\label{rg:sc}
\left(
\dmu 
+ 2 \gamma_m(\alpha_s)
\right)
\Pi^{{\rm S/P}} =  Q^2 \g^{{\rm SS}}_q(\as)
\frac{1}{16\pi^2} 
{}.
\eeq

The (axial) vector and (pseudo)scalar correlators are connected
through  a  Ward identity \cite{david75}
\begin{equation}
q_\mu q_\nu
\Pi^{{\rm V/A}}_{\mu\nu} = 
(m_u \mp m_d)^2 \Pi^{\rm S/P} 
+
(m_u \mp m_d)
(
\langle 
\overline{\psi}_{{\rm q}} \psi_{{\rm q}}
\rangle
\mp
\langle 
\overline{\psi}_{{\rm q'}} \psi_{{\rm q'}}
\rangle 
)
{}\, ,
\label{axial-ward}
\end{equation}
where the vacuum expectation values on the r.h.s.  are
understood within the framework of perturbation theory  and 
the minimal subtractions. 
Equation~(\ref{axial-ward}) leads to the following relation between the
corresponding anomalous dimensions \cite{CheKueKwi92}:
\begin{equation}
\gamma_{{m}}^{\rm VV}
\equiv
-
\gamma_{{q}}^{{\rm SS}}
{}\, .
\label{VV-SS}
\end{equation}
This relation was used in Ref.~\cite{CheKueKwi92}
in order to find  the anomalous dimension
$\gamma_{{m}}^{\rm AA}$ at the $\alpha_s^2$
order starting from
the results of Ref.~\cite{GorKatLarSur90}.

In  what follows we will be interested in quadratic mass corrections
to the  polarization operator $\Pi^{(1)}_{A}$ 
which is  convenient to represent in 
the form (${\bf m} = \{ m_u, m_d, m{} \}$):  
\beq
\Pi^{(1)}_{V/A}( Q^2,{\bf m},\mu,\alpha_s) 
= 
\frac{3}{16\pi^2}\Pi^{(1)}_{V/A,0}(\frac{\mu^2}{Q^2}, \alpha_s)
+
\frac{3}{16\pi^2}\Pi^{(1)}_{V/A,2}(\frac{\mu^2}{Q^2},{\bf m},\alpha_s)
+ {\cal O}({\bf m^4})
{}.
\label{mass-exp}
\eeq
Here  the first term on the rhs  corresponds  to the massless
limit while the second term stands for quadratic mass corrections. 
Note that $\Pi^{(1)}_{V/A,2}$ is a second order polynomial in quark
masses: a logarithmic dependence on quark masses may appear starting
from $m^4$ terms only\footnote{Provided of course that one uses a mass
 independent renormalization scheme like the $\msbar$-scheme employed
 in this work.}.

{}From the RG equation (\ref{rgea})  we arrive at the following 
equation for  $\Pi^{(1)}_{V/A,2}$:
\beq
\dmu \Pi^{(1)}_{V/A,2} = \frac{1}{3}(m_u \mp m_d)^2 \gvvm(\as)
\label{rgPi1}
{}
\eeq
or, equivalently, ($L_q = \ln\frac{\mu^2}{Q^2}$)
\beq
\frac{\prd }{\prd L_q} \Pi^{(1)}_{V/A,2} =
\frac{1}{3}(m_u \mp m_d)
\gvvm 
-\left(
 \beta \as \frac{\prd }{\prd \alpha_s}
+ 2\gamma_m
\right) \Pi^{(1)}_{V/A,2}
\label{rgPi2}
{}.
\eeq

The last relation explicitly demonstrates that 
$R^{V/A}_2$ --- the absorptive part of $\Pi^{(1)}_{V/A,2}$ --- depends  
in order $\alpha_s^n$ on 
the very function $\Pi^{(1)}_{V/A,2}$ which is multiplied by at least
one factor of $\alpha_s$. This  means that one needs to know
$\Pi^{(1)}_{V/A,2}$ up to order $\alpha_s^{n-1}$ only to unambiguously
reconstruct all $Q$-dependent terms in $\Pi^{(1)}_{V/A,2}$ to 
$\alpha_s^n$, provided, of course, the beta function and anomalous
dimensions $\gamma_m$ and $\gvvm$ are known to $\alpha_s^n$.

This observation was made first  in \cite{ChetKuhn90} where it was
used to find the absorptive part $R^{V}_2$ in order $\alpha_s^3$ for the
case of the diagonal vector current (that is for the
case of $m_u = m_d$). In the
present paper we will use the results of a recent calculation of
$\gssq$ \cite{gssq} to order $\alpha_s^3$ to determine the absorptive
part $R^{V/A}_2$ to the same order in the
general case of non-diagonal currents. 
\section{Calculation and results}
The result for the function  
$\Pi^{(2)}_{V/A,2}$
in the general non-diagonal case to order
$\as^2$ was first published in Ref.~\cite{Chetyrkin93}.
On the other hand,  the Ward identity (\ref{axial-ward})
expresses the combination 
$\Pi^{(1)}_{V/A,2}/Q^2  - \Pi^{(2)}_{V/A,2} $
in terms of the {\em massless} polarization operator
$\Pi^S$ known from Refs.~\cite{GorKatLarSur90,Karl94}.
A sum  of these two functions leads us to
the following result for $\Pi^{(1)}_{V,2}$ 
\begin{eqnarray}
\lefteqn{\Pi^{(1)}_{V,2} =
{}\frac{m_{-}^2}{Q^2}
\left[
2 
+2  \lgm\frac{\mu^2}{Q^2}
\right]
{+}\frac{m_{+}^2}{Q^2}
\left[
-2\right
]}
\nonumber\\
&{+}&\frac{m_{-}^2}{Q^2} \frac{\alpha_s}{\pi}
\left[
\frac{107}{6} 
-8  \,\zeta(3)
+\frac{22}{3}  \lgm\frac{\mu^2}{Q^2}
+2  \lgm^2\frac{\mu^2}{Q^2}
\right]
{+}\frac{m_{+}^2}{Q^2} \frac{\alpha_s}{\pi}
\left[
-\frac{16}{3} 
-4  \lgm\frac{\mu^2}{Q^2}
\right]
\nonumber\\
&{+}&\frac{m_{-}^2}{Q^2}\left(\frac{\alpha_s}{\pi}\right)^2
\left[
\frac{3241}{18} 
-129  \,\zeta(3)
-\frac{1}{2}  \,\zeta(4)
+55  \,\zeta(5)
-\frac{857}{108}  \,n_f
+\frac{32}{9}  \,\zeta(3) \,n_f
\Break
\phantom{+\frac{m_{-}^2}{Q^2}\left(\frac{\alpha_s}{\pi}\right)^2}
+\frac{8221}{72}  \lgm\frac{\mu^2}{Q^2}
-39  \,\zeta(3) \lgm\frac{\mu^2}{Q^2}
-\frac{151}{36}  \,n_f \lgm\frac{\mu^2}{Q^2}
+\frac{4}{3}  \,\zeta(3) \,n_f \lgm\frac{\mu^2}{Q^2}
\Break
\phantom{+\frac{m_{-}^2}{Q^2}\left(\frac{\alpha_s}{\pi}\right)^2}
+\frac{155}{6}  \lgm^2\frac{\mu^2}{Q^2}
-\frac{8}{9}  \,n_f \lgm^2\frac{\mu^2}{Q^2}
+\frac{19}{6}  \lgm^3\frac{\mu^2}{Q^2}
-\frac{1}{9}  \,n_f \lgm^3\frac{\mu^2}{Q^2}
\right]
\nonumber\\
&{+}&\frac{m_{+}^2}{Q^2}\left(\frac{\alpha_s}{\pi}\right)^2
\left[
-\frac{19691}{216} 
-\frac{124}{27}  \,\zeta(3)
+\frac{1045}{27}  \,\zeta(5)
+\frac{95}{36}  \,n_f
-\frac{253}{6}  \lgm\frac{\mu^2}{Q^2}
\Break
\phantom{+\frac{m_{+}^2}{Q^2}\left(\frac{\alpha_s}{\pi}\right)^2}
+\frac{13}{9}  \,n_f \lgm\frac{\mu^2}{Q^2}
-\frac{19}{2}  \lgm^2\frac{\mu^2}{Q^2}
+\frac{1}{3}  \,n_f \lgm^2\frac{\mu^2}{Q^2}
\right]
\nonumber\\
&{+}&\frac{m^2}{Q^2}\left(\frac{\alpha_s}{\pi}\right)^2
\left[
\frac{128}{9}  
-\frac{32}{3}  \,\zeta(3) 
\right]
{}.
\label{Pi1V2}
\end{eqnarray}
Here $m_- =
m_u - m_d$ and $m_+ = m_u + m_d$, $Q^2 = -q^2$, 
all masses as well as QCD coupling constant $\alpha_s$ are understood
to be taken at a generic value of the t' Hooft~mass $\mu$. All
correlators are  renormalized within $\msbar$-scheme.  We have also
checked (\ref{Pi1V2})  by a direct calculation with the help of the
program MINCER \cite{mincer2} written for the symbolic manipulation system
FORM \cite{Ver91}. In a particular case of 
$m_u = m_d$ Eq.~(\ref{Pi1V2}) is in
agreement with Refs.~\cite{GorKatLar86,levan94}.

Now, as was shown  in \cite{ChetKuhn90}
the anomalous dimension $\gaam \equiv = -\gssq$, and, thus, from the
results of \cite{gssq} we have:
\begin{eqnarray}
\lefteqn{\gvvm = -\g^{SS}_q =  6\left\{ 
1
{+} \frac{5}{3}\frac{\alpha_s}{\pi}
\right. 
{+}\left(\frac{\alpha_s}{\pi}\right)^2
\left[
\frac{455}{72} 
-\frac{1}{2}  \,\zeta(3)
-\frac{1}{3}  \,n_f
\right]}
\nonumber\\
&{+}&\left(\frac{\alpha_s}{\pi}\right)^3
\left[
\frac{157697}{5184} 
-\frac{1645}{216}  \,\zeta(3)
+\frac{15}{8}  \,\zeta(4)
+\frac{65}{12}  \,\zeta(5)
-\frac{14131}{7776}  \,n_f
\Break
\left.
\phantom{+\left(\frac{\alpha_s}{\pi}\right)^3}
-\frac{13}{9}  \,\zeta(3) \,n_f
-\frac{11}{12}  \,\zeta(4) \,n_f
-\frac{1625}{11664}  \, n_f^2
+\frac{1}{9}  \,\zeta(3) \, n_f^2
\right]
\right\}
{}.
\label{gVVm}
\end{eqnarray}

At last, integrating eq.~(\ref{rgPi2}) we find the spectral density
$R^{V}_2$ in general case  to order $\as^3$:
\beq
R^V_2 = 3\left\{
\frac{m_+^2}{s} r^V_{2,+} 
+
\frac{m_-^2}{s}r^V_{2,-}
+
\frac{m^2}{s}r^V_{2,0}
\right\}
{},
\label{R2V}
\eeq
where the functions $r^V$ are
\begin{eqnarray}
\lefteqn{r_{2,+}^{V}   = 
 3 \frac{\alpha_s}{\pi}
{+} \left(\frac{\alpha_s}{\pi}\right)^2
\left[
\frac{253}{8} 
-\frac{13}{12}  \,n_f 
+\frac{57}{4}  \lgm\frac{\mu^2}{s}
-\frac{1}{2}  \,n_f  \lgm\frac{\mu^2}{s}
\right]}
\nonumber\\
&{+}& \left(\frac{\alpha_s}{\pi}\right)^3
\left[
\frac{1261}{2} 
-\frac{285}{16}  \pi^2
+\frac{155}{6}  \,\zeta(3)
-\frac{5225}{24}  \,\zeta(5)
-\frac{2471}{54}  \,n_f 
+\frac{17}{12}  \pi^2 \,n_f 
\label{r2+ }
\right. \\ &{}& \left.
\phantom{+ \left(\frac{\alpha_s}{\pi}\right)^3}
-\frac{197}{54}  \,\zeta(3) \,n_f 
+\frac{1045}{108}  \,\zeta(5) \,n_f 
+\frac{125}{216}  \, n_f^2
-\frac{1}{36}  \pi^2 \, n_f^2
+\frac{4505}{16}  \lgm\frac{\mu^2}{s}
\Break
\phantom{+ \left(\frac{\alpha_s}{\pi}\right)^3}
-\frac{175}{8}  \,n_f  \lgm\frac{\mu^2}{s}
+\frac{13}{36}  \, n_f^2 \lgm\frac{\mu^2}{s}
+\frac{855}{16}  \lgm^2\frac{\mu^2}{s}
-\frac{17}{4}  \,n_f  \lgm^2\frac{\mu^2}{s}
+\frac{1}{12}  \, n_f^2 \lgm^2\frac{\mu^2}{s}
\right]
{},
\nonumber
\end{eqnarray}
\begin{eqnarray}
\lefteqn{r_{2,-}^{V} =  
-\frac{3}{2}
{+}  \frac{\alpha_s}{\pi}
\left[
-\frac{11}{2} 
-3  \lgm\frac{\mu^2}{s}
\right]
}
\nonumber\\
&{+}& \left(\frac{\alpha_s}{\pi}\right)^2
\left[
-\frac{8221}{96} 
+\frac{19}{8}  \pi^2
+\frac{117}{4}  \,\zeta(3)
+\frac{151}{48}  \,n_f 
-\frac{1}{12}  \pi^2 \,n_f 
\Break
\phantom{+ \left(\frac{\alpha_s}{\pi}\right)^2}
-  \,\zeta(3) \,n_f 
-\frac{155}{4}  \lgm\frac{\mu^2}{s}
+\frac{4}{3}  \,n_f  \lgm\frac{\mu^2}{s}
-\frac{57}{8}  \lgm^2\frac{\mu^2}{s}
+\frac{1}{4}  \,n_f  \lgm^2\frac{\mu^2}{s}
\right]
\nonumber\\
&{+}& \left(\frac{\alpha_s}{\pi}\right)^3
\left[
-\frac{4544045}{3456} 
+\frac{335}{6}  \pi^2
+\frac{118915}{144}  \,\zeta(3)
-\frac{635}{2}  \,\zeta(5)
+\frac{71621}{648}  \,n_f 
\Break
\phantom{+ \left(\frac{\alpha_s}{\pi}\right)^3}
-\frac{209}{48}  \pi^2 \,n_f 
-54  \,\zeta(3) \,n_f 
+\frac{5}{4}  \,\zeta(4) \,n_f 
+\frac{55}{4}  \,\zeta(5) \,n_f 
-\frac{13171}{7776}  \, n_f^2
\Break
\phantom{+ \left(\frac{\alpha_s}{\pi}\right)^3}
+\frac{2}{27}  \pi^2 \, n_f^2
+\frac{13}{18}  \,\zeta(3) \, n_f^2
-\frac{4693}{6}  \lgm\frac{\mu^2}{s}
+\frac{285}{16}  \pi^2 \lgm\frac{\mu^2}{s}
+\frac{1755}{8}  \,\zeta(3) \lgm\frac{\mu^2}{s}
\Break
\phantom{+ \left(\frac{\alpha_s}{\pi}\right)^3}
+\frac{8909}{144}  \,n_f  \lgm\frac{\mu^2}{s}
-\frac{17}{12}  \pi^2 \,n_f  \lgm\frac{\mu^2}{s}
-\frac{59}{4}  \,\zeta(3) \,n_f  \lgm\frac{\mu^2}{s}
-\frac{209}{216}  \, n_f^2 \lgm\frac{\mu^2}{s}
\Break
\phantom{+ \left(\frac{\alpha_s}{\pi}\right)^3}
+\frac{1}{36}  \pi^2 \, n_f^2 \lgm\frac{\mu^2}{s}
+\frac{1}{3}  \,\zeta(3) \, n_f^2 \lgm\frac{\mu^2}{s}
-\frac{335}{2}  \lgm^2\frac{\mu^2}{s}
+\frac{209}{16}  \,n_f  \lgm^2\frac{\mu^2}{s}
\Break
\phantom{+ \left(\frac{\alpha_s}{\pi}\right)^3}
-\frac{2}{9}  \, n_f^2 \lgm^2\frac{\mu^2}{s}
-\frac{285}{16}  \lgm^3\frac{\mu^2}{s}
+\frac{17}{12}  \,n_f  \lgm^3\frac{\mu^2}{s}
-\frac{1}{36}  \, n_f^2 \lgm^3\frac{\mu^2}{s}
\right]
{},
\nonumber\\
\label{r2-}
\end{eqnarray}
\beq
r_{2,0}^{V} = 
\left(\frac{\alpha_s}{\pi}\right)^3
\left[
-80 
+60  \,\zeta(3)
+\frac{32}{9}  \,n_f 
-\frac{8}{3}  \,\zeta(3) \,n_f 
\right]
{}.
\label{r20}
\eeq

The expressions for the hadronic decay rates of the intermediate
bosons read:
\begin{eqnarray}
\G(Z \to \mbox{hadrons} ) = \G^Z_0 
\left[
\right.
 &{}&
\sum_f ((g^f_V)^2 + (g^f_A)^2)\left(R_0(s) 
 + 
R_2^V(s,0,0,\sqrt{m^2_b+m_c^2}\right)
\nonumber
\\
&+&\sum_{f=b,c} (g^f_V)^2 R_2^V(s,m_f,m_f,0)
\nonumber
\\
&+& 
\left.
\sum_{f=b,c}(g^f_A)^2 R_2^V(s,m_f,m_f,0)
\right]
\label{Zff}
{},
\end{eqnarray}
\begin{eqnarray}
\G(W \to \mbox{hadrons}) = \G^W_0 
\left[
\right.
 &{}&
 2 \left( R_0(s) +R_2^{V}(s,0,0,\sqrt{m^2_b+m_c^2}) ) \right)
\label{Wff}
\\
&+& \frac{1}{2}\sum_{\dsp {i=u,c}\atop{\dsp j=d,s,b}} |V_{i,j}|^2( R_2^V(s,m_{i},m_{j},0) 
+  
R_2^A(s,m_{i},m_{j},0) )
\left.
\right]
\nonumber
{},
\end{eqnarray}
with $\G_0 = \frac{G_F M_{Z/W}^3}{6\pi \sqrt{2}}$,
$g^{{f}}_V =  I^{{f}}_3 - 2Q_{{f}}\sin^2 \theta_{{\rm w}},
 \ \ 
 g^{{f}}_A =  I^{{f}}_3$
 and
$V_{ij}$ being the CKM matrix.

Here $R_0(s)$ is the (non-singlet part) of the ratio $R(s)$ in 
massless QCD; it was  computed to $\as^3$ in
\cite{GorKatLar91,SurSam91} and confirmed in \cite{gluino};
it reads:
\begin{equation}\label{apxc2}
\begin{array}{rl}\displaystyle
R_0(s) = 
& \displaystyle
 3
\Bigg\{
1+\frac{\alpha_s}{\pi}
\\ & \displaystyle
+\left(\frac{\alpha_s}{\pi}\right)^2
\left[
\frac{365}{24}-11\zeta(3)
+n_f\left(
          -\frac{11}{12}+\frac{2}{3}\zeta(3)
    \right)
+\left(
-\frac{11}{4}+\frac{1}{6}n_f
\right)\ln\,\frac{s}{\mu^2}
\right]
\\ & \displaystyle
+\left(\frac{\alpha_s}{\pi}\right)^3
\Bigg[
\frac{87029}{288}-\frac{121}{48}\pi^2
   -\frac{1103}{4}\zeta(3) +\frac{275}{6}\zeta(5)
\\ & \displaystyle
\hphantom{\left(\frac{}{}\right)^2}
+n_f\left(
  -\frac{7847}{216}+\frac{11}{36}\pi^2
  +\frac{262}{9}\zeta(3)-\frac{25}{9}\zeta(5)
    \right)
+n_f^2\left(
  \frac{151}{162}-\frac{1}{108}\pi^2
  -\frac{19}{27}\zeta(3)
     \right)
\\ & \displaystyle
\hphantom{\left(\frac{}{}\right)^2}
+\left(
  -\frac{4321}{48}+\frac{121}{2}\zeta(3)
  +n_f\left[
       \frac{785}{72}-\frac{22}{3}\zeta(3)
      \right]
  +n_f^2\left[
       -\frac{11}{36}+\frac{2}{9}\zeta(3)
        \right]
\right)\ln\,\frac{s}{\mu^2}W
\\ & \displaystyle
\hphantom{\left(\frac{}{}\right)^2}
+\left(
  \frac{121}{16}-\frac{11}{12}n_f
   +\frac{1}{36}n_f^2
\right)\ln^2\frac{s}{\mu^2}
\Bigg]
\Bigg\}
{}.
\end{array}
\end{equation}
In deriving Eqs.~(\ref{Zff}) and (\ref{Wff}) we have assumed 
that 
(i) the top quark is completely   decoupled (the power suppressed 
corrections to this approximation start from the order 
$\frac{s}{m_t^2}\as^2$ 
and  have been studied in Refs.~\cite{Kniehl90,me93,Soper94});  
(ii) all other quarks except for the charmed and bottom ones 
are massless. 
Note that for the case of {\em diagonal} currents
there exist also so-called singlet contributions to $R(s)$.
We will ignore these contributions in what follows  as they
are absent for the case of non-diagonal currents relevant for
the $W$-decay (a detailed discussion of the $Z$-decay rate
including singlet contributions can  be found in \cite{review}).

Taking  into account 
the peculiar structure of the general result (\ref{R2V}),
the last formula  can be written in  a simpler form, viz. 
\begin{eqnarray}
\G(W \to \mbox{hadrons}) = \G^W_0 
\left[
\right.
 &{}&
 2 \left( R_0(s) +R_2^V(s,0,0,\sqrt{m^2_b+m_c^2}) ) \right)
\nonumber
\\
&+&  R_2^V(s,m_{eff},0,0) 
\left.
\right]
\label{Wff2}
{}.
\end{eqnarray}
Here
\[
m_{eff}^2 = \sum_{\dsp {i=u,c}\atop{\dsp j=d,s,b}} |V_{i,j}|^2 (m_i^2 + m_j^2)
\]
and we have taken into account the fact  that 
\[
R^V(s,m_i,m_j,0) + R^A(s,m_i,m_j,0) =  
2 R^V(s,\sqrt{m^2_i+m^2_j},0,0) = 2 R^A(s,\sqrt{m^2_i+m^2_j},0,0)
\]
As a  direct consequence of Eqs.~(2,\ref{R2V})
we obtain the following expressions for particular functions entering into 
(\ref{Zff},\ref{Wff}) 
\begin{eqnarray}
R_2^{V}(s,m,m,0) &=&  \frac{4m^2}{s} 3 r^V_{2,+},
\label{part1}
\\
R_2^{A}(s,m,m,0) &=&  \frac{4m^2}{s} 3 r^V_{2,-},
\\
\label{part2}
R_2^{V}(s,m,0,0) &=& R_2^{A}(s,m,0,0)= \frac{m^2}{s} 3
( r^V_{2,+}+ r^V_{2,-} ),
\\
R_2^{V}(s,0,0,m) &=& R_2^{A}(s,0,0,m) = \frac{m^2}{s}3r_{2,0}
{}.
\label{part3}
\end{eqnarray}
At last, with $n_f=5$ 
and  $\mu^2= s$ the above formulas are simplified to
\begin{eqnarray}
R_2^{V}(s,m,m,0) = 
&{}&\frac{m^2}{s}3\left\{ 
12\frac{\alpha_s}{\pi}
{+}\frac{629}{6}
\left(\frac{\alpha_s}{\pi}\right)^2
\right.
\label{R2Vmm0nf5}
\\
&{+}&
\left.
\left(\frac{\alpha_s}{\pi}\right)^3
\left[
\frac{89893}{54} 
-\frac{1645}{36}  \pi^2
+\frac{820}{27}  \,\zeta(3)
-\frac{36575}{54}  \,\zeta(5)
\right]
\right\}
\nonumber
\end{eqnarray}
\begin{eqnarray}
&{}&\!\!\!\!\!\!\!\!\!R_2^{A}(s,m,m,0) = 
{}\frac{m^2}{s}
3\left\{
-6
-22 \frac{\alpha_s}{\pi}
+\left(\frac{\alpha_s}{\pi}\right)^2
\left[
-\frac{2237}{8} 
+\frac{47}{6}  \pi^2
+97  \,\zeta(3)
\right]
\right.
\nonumber\\
&{+}&\left(\frac{\alpha_s}{\pi}\right)^3
\left.
\left[
-\frac{25024465}{7776} 
+\frac{15515}{108}  \pi^2
+\frac{27545}{12}  \,\zeta(3)
+25  \,\zeta(4)
-995  \,\zeta(5)
\right]
\right\}
\label{R2Amm0nf5}
\end{eqnarray}
\begin{eqnarray}
&{}&R_2^{V}(s,m,0,0)=
\frac{m^2}{s}
3\left\{
-\frac{3}{2}
-\frac{5}{2} \frac{\alpha_s}{\pi}
{+}\left(\frac{\alpha_s}{\pi}\right)^2
\left[
-\frac{4195}{96} 
+\frac{47}{24}  \pi^2
+\frac{97}{4}  \,\zeta(3)
\right]
\right.
\nonumber\\
&{+}&\left(\frac{\alpha_s}{\pi}\right)^3
\left.
\left[
-\frac{12079873}{31104} 
+\frac{2645}{108}  \pi^2
+\frac{251185}{432}  \,\zeta(3)
+\frac{25}{4}  \,\zeta(4)
-\frac{90305}{216}  \,\zeta(5)
\right]
\right\}
\EQN{ }
\end{eqnarray}
or, in the numerical form,
\beq
R_2^{V}(s,m,m,0) = 
\frac{m^2}{s}
3\left\{
 12
\frac{\alpha_s}{\pi}
{+}104.833\left(\frac{\alpha_s}{\pi}\right)^2
{+}547.879\left(\frac{\alpha_s}{\pi}\right)^3
\right\}
\EQN{R2Vmm0NN}
{},
\eeq
\beq
R_2^{A}(s,m,m,0) =  
\frac{m_{}^2}{s}
3\left\{
-6
-22 \frac{\alpha_s}{\pi}
-85.7136\left(\frac{\alpha_s}{\pi}\right)^2
-45.7886 \left(\frac{\alpha_s}{\pi}\right)^3
\right\}
{},
\EQN{R2Amm0NN}
\eeq
\beq
R_2^{V}(s,m,0,0)  \equiv R_2^{A}(s,m,0,0) = 
\frac{m_{}^2}{s}
3\left\{
-1.5
-2.5 \frac{\alpha_s}{\pi}
+ 4.7799\left(\frac{\alpha_s}{\pi}\right)^2
+ 125.523\left(\frac{\alpha_s}{\pi}\right)^3
\right\}
{}.
\EQN{R2Vm00NN}
\eeq
\section{ Acknowledgments}

\noindent
This work was supported by BMFT under Contract 057KA92P(0) 
and INTAS under Contract INTAS-93-0744.
\sloppy
\raggedright
\def\app#1#2#3{{\it Act. Phys. Pol. }{\bf B #1} (#2) #3}
\def\apa#1#2#3{{\it Act. Phys. Austr.}{\bf #1} (#2) #3}
\def\lhc{Proc. LHC Workshop, CERN 90-10}
\def\npb#1#2#3{{\it Nucl. Phys. }{\bf B #1} (#2) #3}
\def\plb#1#2#3{{\it Phys. Lett. }{\bf B #1} (#2) #3}
\def\prd#1#2#3{{\it Phys. Rev. }{\bf D #1} (#2) #3}
\def\pR#1#2#3{{\it Phys. Rev. }{\bf #1} (#2) #3}
\def\prl#1#2#3{{\it Phys. Rev. Lett. }{\bf #1} (#2) #3}
\def\prc#1#2#3{{\it Phys. Reports }{\bf #1} (#2) #3}
\def\cpc#1#2#3{{\it Comp. Phys. Commun. }{\bf #1} (#2) #3}
\def\nim#1#2#3{{\it Nucl. Inst. Meth. }{\bf #1} (#2) #3}
\def\pr#1#2#3{{\it Phys. Reports }{\bf #1} (#2) #3}
\def\sovnp#1#2#3{{\it Sov. J. Nucl. Phys. }{\bf #1} (#2) #3}
\def\jl#1#2#3{{\it JETP Lett. }{\bf #1} (#2) #3}
\def\jet#1#2#3{{\it JETP Lett. }{\bf #1} (#2) #3}
\def\zpc#1#2#3{{\it Z. Phys. }{\bf C #1} (#2) #3}
\def\ptp#1#2#3{{\it Prog.~Theor.~Phys.~}{\bf #1} (#2) #3}
\def\nca#1#2#3{{\it Nouvo~Cim.~}{\bf #1A} (#2) #3}
\def\mpl#1#2#3{{\it Mod. Phys. Lett.~}{\bf A #1} (#2) #3}


\end{document}